# dMotifGreedy: a novel tool for de novo discovery of DNA motifs with enhanced power of reporting distinct motifs

Yupeng Wang[1,2,*], Xinyu Liu[1,2], Michael Kelley[1] and Romdhane Rekaya[1,2,3]

[1]Department of Animal and Dairy Science, University of Georgia, Athens, GA 30602.
[2]Institute of Bioinformatics, University of Georgia, Athens, GA 30602.
[3]Department of Statistics, University of Georgia, Athens, GA 30602.

*Corresponding author

E-mails: YW: wyp1125@uga.edu
          XL: xinyu81@uga.edu
          MK: mck@uga.edu
          RR: rrekaya@uga.edu

# Abstract

De novo discovery of over-represented DNA motifs is one of the major challenges in computational biology. Although numerous tools have been available for de novo motif discovery, many of these tools are subject to local optima phenomena, which may hinder detection of multiple distinct motifs. A greedy algorithm based tool named dMotifGreedy was developed. dMotifGreedy begins by searching for candidate motifs from pair-wise local alignments of input sequences and then computes an optimal global solution for each candidate motif through a greedy algorithm. dMotifGreedy has competitive performance in detecting a true motif and greatly enhanced performance in detecting multiple distinct true motifs. dMotifGreedy is freely available via a stand-alone program at http://lambchop.ads.uga.edu/dmotifgreedy/download.php.

## Introduction

DNA motifs, such as transcription factor binding sites, play key roles in gene expression regulation. As experimental endeavors towards understanding gene expression regulation are laborious and expensive, de novo discovery of DNA motifs is one of the most important problems in bioinformatics. Over one hundred algorithms have been published for motif discovery [1]. However, no algorithm can perfectly resolve the problem of motif discovery [2, 3]. The solution of this problem often relies on some optimization techniques such as greedy algorithm [4], Gibbs sampling [5] and Expectation Maximization [6]. However, these methods are often subject to potential local optima phenomena [2], which may hinder the detection of another motif when one motif has been identified. Although many tools provide the option of reporting multiple motifs, it is often observed that most of the reported motifs are often the adjustments of the same motif (e.g. different widths of a motif or partial change of the binding sites of a motif).

To overcome the local optima phenomena in many of the current strategies for de novo motif discovery, we propose the use of dMotifGreedy, a novel tool for de novo detection of DNA motifs. dMotifGreedy is a two-stage approach. The first stage is pair-wise local alignment of the input sequences, through which top local motifs are selected as candidate motifs. The second stage is a greedy strategy, in which each candidate motif is investigated for its global solution. dMotifGreedy attempts to select many initial local conserved regions and adopts a procedure to merge the same output motifs, giving it improved power in reporting distinct motifs.

## Methods

In this study, L denotes the sequence length; n ($\geq 2$) denotes the number of input genes; w denotes the motif width. The starting point is a set of promoter sequences $S = \{s_1, s_2, ..., s_n\}$. dMotifGreedy algorithm first finds candidate motifs from the $C_n^2$ pair-wise local alignments in S, where the top motifs of a pair-wise alignment that satisfy the motif width constraints and the percentage identity are considered as candidate motifs (the detailed procedure is described in Appendix A). Next, dMotifGreedy algorithm attempts to search for an optimal global solution (that is, a motif that exists in all promoter sequences) for each candidate motif through n-2 iterations, at each of which the best position in one of the remaining sequences that maximizes the information content (IC) of the candidate motif is added. After the optimal global solutions of all candidate motifs are computed, the p-value of each candidate motif is calculated, following the procedure presented by Hertz and Stormo [4]. Finally, duplicated predictions are merged, and the distinct candidate motifs with a p-value smaller than the user-defined threshold are output as results in ascending order of p-values. As in practice, a motif may not exist in all input sequences; for a reported motif, its optimal motif in the search path that minimizes its p-value may be also reported. The time complexity of dMotifGreedy algorithm is $o(L^3 n^4)$. A pseudocode of dMotifGreedy algorithm is shown in Appendix B.

## Results

**Performance and testing**

To systematically evaluate the performance of dMotifGreedy, we compared dMotifGreedy with three popular motif-finding tools that have a multiple motif output: MEME [6], WCONSENSUS [4], and Weeder [7]. Different tools were evaluated in terms of the ability to predict distinct motifs. Thus, duplicate predictions of the same motif were not regarded as an increase of power. In each testing dataset, at least two motifs were placed. To make different tools comparable, we considered the top 10 predictions of each tool for detecting 2 distinct motifs or the top 15 predictions of each tool for detecting 3 distinct motifs.

Evaluation of motif detection tools on their power to predict true motifs over a series of mutation rates of the motifs has been proven to be a good simulation strategy [8]. We first generated a number of datasets containing 2 or 3 distinct motifs that were placed into randomly chosen locations in each of a set of randomly generated background sequences after these distinct motifs were point-mutated, according to given mutation rates (denoted by mr). At least 2 motifs were placed in each dataset with difference in mutation rates (denoted by $\Delta mr$) being 0, 0.05 or 0.10. The average mutation rates ranged from 0.05 to 0.25 with increments of 0.05. The motif widths ranged from 8nt to 16nt. The number of combinations of parameter settings was 84. Each dataset contained 10 sequences with 200nt length. For each parameter setting, we simulated 100 datasets.

We compared the prediction accuracies of different tools at each parameter setting. For each tool, the successful identification of a true motif consisted of the predicted motif consensus overlapping with the true consensus (those that have $\geq 75\%$ matches with the true consensus). The prediction accuracy is defined as the percentage of the successful identification of the true motifs averaged over 100 datasets, a similar definition as in Li et al. [8]. In this study, the prediction accuracy has two levels: 1) successful identification of at least 1 motif, an indicator of the power of detecting a true motif, and 2) successful identification of both/all motifs, an indicator of being less affected by local optima phenomena. The comparison is summarized in Table 1, indicating that dMotifGreedy places second in detecting a true motif and is the best in detecting distinct true motifs. To test the scalability and robustness of different tools, we compared different tools on the datasets with longer sequences or 30% additional random sequences. The results indicate that dMotifGreedy remains the best one in detecting distinct motifs, though the prediction accuracy decreases slightly (Table 2 and Table 3).

To further test the performance of dMotifGreedy on more practical data, we downloaded true motif binding sites of Escherichia coli K12, humans and Drosophila from RegulonDB [9] and JASPAR [10]. True motifs were paired and placed in genomic upstream sequences. The comparison of different tools is summarized in Table 4. The results suggest that dMotifGreedy is always the second best in detecting a true motif and the best in detecting both distinct true motifs.

**Implementation and Availability**

dMotifGreedy is freely available via a stand-alone program at http://lambchop.ads.uga.edu/dmotifgreedy/download.php.

Input: Users need to enter the upstream sequences in FASTA format, number of top motifs in pair-wise local alignment (optional, default is 2), p-value threshold (optional, default is 0.01), minimum motif width (optional, default is 6), maximum motif width (optional, default is 18) and threshold of the percentage identity (optional, default is 0.65).

Output: Putative motifs are shown in a compact format in ascending order of p-values. For each motif, the consensus sequence, IC, p-value and locations of binding sites are displayed.

## Discussion

Our computational analysis of multiple datasets indicates that dMotifGreedy is not as fast as MEME or WCONSENSUS but has a comparable running time with Weeder. The running time increases polynomially with n and L. dMotifGreedy is efficient in small datasets (say, n<10 and L<500), which can be completed within several minutes.

## Appendix A

**1) Selecting top motifs from pair-wise local alignment**
Local alignment is performed based on the Smith-Waterman algorithm without gaps. The matrix H is initialized as follows:

$$\begin{cases} H(i,0) = 0, & 0 \le i \le u \\ H(0, j) = 0, & 0 \le j \le v \\ H(i, j) = \begin{cases} \max(0, H(i, j) - 1), & \text{if mismatch} \\ H(i, j) + 1, & \text{if match} \end{cases} \end{cases}$$

where u and v are the sequence lengths.

Another matrix T, denoting which positions are involved in motifs, is initialized as follows:

$$T(i, j) = 0, \quad 1 \le i \le u, 1 \le j \le v.$$

For the first motif, the end position is determined by:

$$\arg\max_{1 \le i \le u, 1 \le j \le v}(H(i, j)),$$

and the start position is chosen as the nearest "1" in front of the end position in the diagonal passing the end position.

When a motif is selected, T is updated as:

$$T(i, j) = \begin{cases} 1, & \text{if } (i,j) \text{ involved in a motif} \\ 0, & \text{otherwise} \end{cases}.$$

Then, H is updated as:

$$H(i, j) = \begin{cases} 0, & \text{if } T(i,j)=1 \\ \begin{cases} \max(0, H(i, j) - 1), & \text{if mismatch} \\ H(i, j) + 1, & \text{if match} \end{cases}, & \text{if } T(i,j)=0 \end{cases}$$

The subsequent motifs are selected following the rules described above.

**2) Selecting candidate motifs from the top motifs of a local alignment**

To be selected as a candidate, a local motif from the local alignment should satisfy some constraints. The first constraints are minimum and maximum motif widths, which may be specified by the user. The default values are $6 \leq w \leq 18$. The second constraint is that the percentage identity (PI) determined by the H value at the end position divided by the motif width w ($0 \leq PI \leq 1$), should be greater than a user-defined threshold. The default threshold is 0.65.

**3) Information content (IC)**

The information content (IC) of a motif m is defined as:

$$IC_m = \sum_{j=1}^{w} \sum_{i=1}^{A} f_{i,j} \ln \frac{f_{i,j}}{p_i},$$

where A is the alphabet of nucleotides (A, C, G, T), $f_{i,j}$ is the frequency that letter i occurs at location j, and $p_i$ is the prior probability of letter i.

**4) Determination of the optimal motif in the search path**

As mentioned in the paper, a candidate motif generated from the local alignment stage, searches for its global solution (a motif that exists in all promoter sequences) through sequentially adding a position that maximizes its IC from one of the remaining sequences. However, a true motif may not exist in all input sequences. Thus, an optimal motif in the search path should also be reported. The optimal motif in the search path consists of the binding sites that decrease its p-value (i.e., the binding site that increases its p-value and the subsequent binding sites are discarded).

**5) Merging the same candidate motifs**

The candidate motifs are sorted in ascending order of p-values. If the consensus of a lower ranking candidate motif completely overlaps with that of a higher ranking one (that is, without any mismatch), only the higher ranking one is reported.

# Appendix B

**The pseudo code of dMotifGreedy algorithm**

**dMotifGreedy algorithm**

**Input**
$S$: $n$ upstream DNA sequences of length $L$
$x$: number of top motifs in a pair-wise comparison of the initial stage:
$p$: p-value threshold, the default value is 0.01
$minw$, $maxw$: minimum and maximum motif widths
$c$: percentage identity cutoff

**Output**
$M$: a set of DNA motifs with p-value $< p$

/\***Phase 1: search for candidate motifs from pair-wise local alignment of the input sequences**\*/
for $i=1$ to $n-1$
  for $j=i+1$ to n
    local alignment (sequence $i$, sequence $j$)
    find top $x$ local motifs
    The local motifs with $minw \leq w \leq maxw$ and the percentage identity (PI) $> c$ are
    selected as candidate motifs and added to $M$ (their locations are recorded)
  end for
end for

/\***Phase 2: search for an optimal global solution for each candidate motif**\*/
for each candidate motif $m$ in $M$
  for $k=1$ to $n-2$
    search for the position in the remaining sequences that maximizes the IC of the
    candidate motif, and add the new location into the record of the candidate motif
  end for
end for

/\***Phase 3: screen candidate motifs**\*/
for each candidate motif $m$ in $M$
  compute the p-value of $m$
end for
sort $M$ in ascending order of p-values and merge the same motifs
Search for optimal motifs in the search paths
delete the candidate motifs with p-value $\geq p$ from $M$
return $M$

# Tables

**Table 1. Comparison of prediction accuracy between MEME (M), WCONSENSUS (C), Weeder (W) and dMotifGreedy (D) with respect to different combinations of widths of distinct motifs.**

| w | Δmr | Average mr | Prediction accuracy ||||||||
| | | | ≥1 motif identified |||| All motifs identified ||||
| | | | M | C | W | D | M | C | W | D |
|---|---|---|---|---|---|---|---|---|---|---|
| 8, 8 | 0 | 0.05 | 1 | 1 | 1 | 1 | 0.91 | 0.79 | 1 | 1 |
| | | 0.10 | 0.99 | 1 | 1 | 1 | 0.76 | 0.49 | 0.96 | 1 |
| | | 0.15 | 0.98 | 1 | 1 | 1 | 0.49 | 0.33 | 0.66 | 0.88 |
| | | 0.20 | 0.88 | 1 | 0.84 | 1 | 0.42 | 0.22 | 0.25 | 0.74 |
| | | 0.25 | 0.78 | 0.96 | 0.61 | 0.90 | 0.21 | 0.25 | 0.07 | 0.38 |
| | 0.5 | 0.05 | 1 | 1 | 1 | 1 | 0.87 | 0.31 | 0.98 | 1 |
| | | 0.10 | 1 | 1 | 1 | 1 | 0.69 | 0.34 | 0.94 | 0.96 |
| | | 0.15 | 0.97 | 1 | 0.99 | 1 | 0.56 | 0.18 | 0.72 | 0.81 |
| | | 0.20 | 0.94 | 0.99 | 0.89 | 0.98 | 0.41 | 0.15 | 0.36 | 0.58 |
| | | 0.25 | 0.78 | 0.97 | 0.60 | 0.87 | 0.23 | 0.18 | 0.15 | 0.37 |
| | 1.0 | 0.10 | 0.99 | 1 | 1 | 1 | 0.64 | 0.02 | 0.69 | 0.97 |
| | | 0.15 | 0.97 | 1 | 1 | 1 | 0.61 | 0.07 | 0.54 | 0.81 |
| | | 0.20 | 0.92 | 0.99 | 0.92 | 0.98 | 0.35 | 0.11 | 0.29 | 0.55 |
| | | 0.25 | 0.89 | 0.99 | 0.71 | 0.96 | 0.31 | 0.13 | 0.11 | 0.41 |
| 10, 10 | 0 | 0.05 | 1 | 1 | 1 | 1 | 0.99 | 0.79 | 0.96 | 1 |
| | | 0.10 | 0.97 | 1 | 1 | 1 | 0.77 | 0.49 | 0.84 | 1 |
| | | 0.15 | 0.94 | 1 | 1 | 0.99 | 0.60 | 0.31 | 0.63 | 0.97 |
| | | 0.20 | 0.79 | 1 | 0.89 | 0.99 | 0.28 | 0.22 | 0.31 | 0.82 |
| | | 0.25 | 0.56 | 0.92 | 0.57 | 0.93 | 0.12 | 0.18 | 0.04 | 0.41 |
| | 0.05 | 0.05 | 1 | 1 | 1 | 1 | 0.98 | 0.31 | 0.99 | 1 |
| | | 0.10 | 1 | 1 | 1 | 1 | 0.83 | 0.34 | 0.93 | 0.99 |
| | | 0.15 | 0.94 | 1 | 0.99 | 1 | 0.57 | 0.15 | 0.69 | 0.94 |
| | | 0.20 | 0.78 | 0.99 | 0.90 | 0.99 | 0.27 | 0.12 | 0.31 | 0.73 |
| | | 0.25 | 0.61 | 0.95 | 0.61 | 0.95 | 0.14 | 0.10 | 0.08 | 0.42 |
| | 0.10 | 0.10 | 1 | 1 | 1 | 1 | 0.82 | 0.02 | 0.56 | 0.97 |
| | | 0.15 | 0.90 | 1 | 1 | 1 | 0.53 | 0.07 | 0.41 | 0.90 |
| | | 0.20 | 0.88 | 0.98 | 0.95 | 0.99 | 0.25 | 0.08 | 0.29 | 0.67 |
| | | 0.25 | 0.51 | 0.95 | 0.67 | 0.96 | 0.07 | 0.08 | 0.04 | 0.26 |
| 12, 12 | 0 | 0.05 | 1 | 1 | 1 | 1 | 1 | 0.90 | 0.99 | 1 |
| | | 0.10 | 1 | 1 | 1 | 1 | 0.91 | 0.54 | 0.87 | 1 |
| | | 0.15 | 0.92 | 1 | 1 | 1 | 0.63 | 0.35 | 0.77 | 1 |
| | | 0.20 | 0.79 | 1 | 1 | 1 | 0.32 | 0.34 | 0.62 | 0.95 |
| | | 0.25 | 0.60 | 0.99 | 0.93 | 0.98 | 0.13 | 0.24 | 0.36 | 0.59 |
| | 0.05 | 0.05 | 1 | 1 | 1 | 1 | 0.99 | 0.37 | 0.94 | 1 |
| | | 0.10 | 1 | 1 | 1 | 1 | 0.88 | 0.30 | 0.72 | 1 |
| | | 0.15 | 0.98 | 1 | 1 | 1 | 0.58 | 0.27 | 0.69 | 0.96 |
| | | 0.20 | 0.88 | 1 | 1 | 1 | 0.28 | 0.25 | 0.51 | 0.89 |
| | | 0.25 | 0.61 | 0.98 | 0.92 | 0.98 | 0.09 | 0.15 | 0.31 | 0.60 |
| | 0.10 | 0.10 | 1 | 1 | 1 | 1 | 0.78 | 0.01 | 0.18 | 1 |
| | | 0.15 | 1 | 1 | 1 | 1 | 0.51 | 0.07 | 0.30 | 0.94 |
| | | 0.20 | 0.91 | 1 | 1 | 1 | 0.21 | 0.05 | 0.19 | 0.76 |
| | | 0.25 | 0.72 | 1 | 0.96 | 0.99 | 0.05 | 0.11 | 0.16 | 0.42 |
| 16, 16 | 0 | 0.05 | 1 | 1 | - | 1 | 1 | 0.88 | - | 1 |
| | | 0.10 | 1 | 1 | - | 1 | 1 | 0.76 | - | 1 |
| | | 0.15 | 0.99 | 1 | - | 1 | 0.93 | 0.46 | - | 0.99 |
| | | 0.20 | 0.94 | 1 | - | 1 | 0.53 | 0.38 | - | 0.89 |
| | | 0.25 | 0.72 | 1 | - | 0.98 | 0.21 | 0.25 | - | 0.56 |
| | 0.05 | 0.05 | 1 | 1 | - | 1 | 1 | 0.38 | - | 1 |
| | | 0.10 | 1 | 1 | - | 1 | 1 | 0.40 | - | 1 |
| | | 0.15 | 1 | 1 | - | 1 | 0.88 | 0.31 | - | 0.97 |
| | | 0.20 | 0.99 | 1 | - | 0.99 | 0.63 | 0.20 | - | 0.84 |
| | | 0.25 | 0.67 | 1 | - | 0.97 | 0.18 | 0.19 | - | 0.49 |
| | 0.10 | 0.10 | 1 | 1 | - | 1 | 0.95 | 0.03 | - | 0.99 |
| | | 0.15 | 1 | 1 | - | 1 | 0.79 | 0.08 | - | 0.93 |
| | | 0.20 | 0.96 | 1 | - | 1 | 0.45 | 0.05 | - | 0.74 |
| | | 0.25 | 0.83 | 1 | - | 0.98 | 0.20 | 0.05 | - | 0.39 |
| 8, 12 | 0 | 0.05 | 1 | 1 | 1 | 1 | 0.98 | 0.06 | 0.94 | 1 |
| | | 0.10 | 1 | 1 | 1 | 1 | 0.87 | 0.10 | 0.96 | 0.99 |
| | | 0.15 | 0.94 | 1 | 1 | 1 | 0.69 | 0.12 | 0.80 | 0.94 |
| | | 0.20 | 0.87 | 0.99 | 0.99 | 0.98 | 0.43 | 0.12 | 0.68 | 0.73 |
| | | 0.25 | 0.72 | 0.96 | 0.97 | 0.98 | 0.17 | 0.22 | 0.37 | 0.57 |
| | 0.05 | 0.05 | 1 | 1 | 1 | 1 | 0.99 | 0.15 | 0.79 | 1 |
| | | 0.10 | 1 | 1 | 1 | 1 | 0.94 | 0.12 | 0.84 | 1 |
| | | 0.15 | 0.97 | 1 | 1 | 1 | 0.66 | 0.09 | 0.70 | 0.91 |
| | | 0.20 | 0.88 | 1 | 0.99 | 1 | 0.37 | 0.11 | 0.62 | 0.75 |
| | | 0.25 | 0.72 | 0.94 | 0.95 | 0.98 | 0.14 | 0.22 | 0.38 | 0.52 |
| | 0.10 | 0.10 | 1 | 1 | 1 | 1 | 0.83 | 0.33 | 0.62 | 1 |
| | | 0.15 | 1 | 1 | 1 | 1 | 0.63 | 0.19 | 0.62 | 0.86 |
| | | 0.20 | 0.90 | 1 | 1 | 1 | 0.39 | 0.11 | 0.39 | 0.67 |
| | | 0.25 | 0.82 | 0.97 | 0.97 | 0.97 | 0.21 | 0.19 | 0.40 | 0.48 |
| 8, 10, 12 | 0 | 0.05 | 1 | 1 | 1 | 1 | 0.98 | 0.05 | 0.43 | 1 |
| | | 0.10 | 1 | 1 | 1 | 1 | 0.88 | 0.04 | 0.38 | 0.99 |
| | | 0.15 | 1 | 1 | 1 | 1 | 0.55 | 0.07 | 0.20 | 0.89 |
| | | 0.20 | 0.96 | 1 | 1 | 1 | 0.28 | 0.01 | 0.03 | 0.60 |
| | | 0.25 | 0.85 | 0.98 | 0.93 | 0.99 | 0.10 | 0.05 | 0.03 | 0.33 |
| | 0.5 | 0.05 | 1 | 1 | 1 | 1 | 0.99 | 0.09 | 0.30 | 1 |
| | | 0.10 | 1 | 1 | 1 | 1 | 0.87 | 0.04 | 0.23 | 0.97 |
| | | 0.15 | 1 | 1 | 1 | 1 | 0.57 | 0.05 | 0.20 | 0.85 |
| | | 0.20 | 0.96 | 0.99 | 1 | 1 | 0.21 | 0.02 | 0.09 | 0.52 |
| | | 0.25 | 0.85 | 0.96 | 0.94 | 0.99 | 0.10 | 0.07 | 0.03 | 0.26 |
| | 1.0 | 0.10 | 1 | 1 | 1 | 1 | 0.88 | 0.02 | 0.28 | 0.94 |
| | | 0.15 | 1 | 1 | 1 | 1 | 0.55 | 0.02 | 0.21 | 0.72 |
| | | 0.20 | 0.96 | 1 | 1 | 0.99 | 0.28 | 0.05 | 0.07 | 0.48 |
| | | 0.25 | 0.85 | 0.97 | 0.97 | 1 | 0.10 | 0.04 | 0.02 | 0.26 |

*w*: motif width, *mr*: mutation rate, and $\Delta mr$: difference in mutation rates

**Table 2. Comparison of prediction accuracy between MEME (M), WCONSENSUS (C), Weeder (W) and dMotifGreedy (D) on datasets with 2 embedded motifs of 12nt and longer sequence lengths.**

| L | $\Delta mr$ | Average *mr* | Prediction accuracy | | | | | | | |
|---|---|---|---|---|---|---|---|---|---|---|
| | | | ≥1 motif identified | | | | 2 motifs identified | | | |
| | | | M | C | W | D | M | C | W | D |
| 400 | 0 | 0.05 | 1 | 1 | 1 | 1 | 1 | 0.81 | 1 | 1 |
| | | 0.10 | 0.97 | 0.99 | 0.99 | 1 | 0.90 | 0.51 | 0.82 | 1 |
| | | 0.15 | 0.88 | 1 | 1 | 1 | 0.58 | 0.38 | 0.77 | 0.94 |
| | | 0.20 | 0.71 | 1 | 0.90 | 0.97 | 0.28 | 0.23 | 0.39 | 0.74 |
| | | 0.25 | 0.38 | 0.89 | 0.69 | 0.93 | 0.05 | 0.13 | 0.15 | 0.41 |
| | 0.05 | 0.05 | 1 | 1 | 1 | 1 | 0.98 | 0.29 | 0.89 | 1 |
| | | 0.10 | 1 | 1 | 1 | 1 | 0.80 | 0.25 | 0.75 | 1 |
| | | 0.15 | 0.92 | 1 | 1 | 1 | 0.47 | 0.27 | 0.61 | 0.96 |
| | | 0.20 | 0.72 | 1 | 1 | 1 | 0.13 | 0.15 | 0.38 | 0.72 |
| | | 0.25 | 0.43 | 0.92 | 0.68 | 0.92 | 0.01 | 0.07 | 0.09 | 0.32 |
| | 0.10 | 0.10 | 1 | 1 | 1 | 1 | 0.78 | 0.07 | 0.27 | 0.97 |
| | | 0.15 | 0.97 | 1 | 1 | 1 | 0.42 | 0.02 | 0.23 | 0.82 |
| | | 0.20 | 0.86 | 1 | 1 | 1 | 0.15 | 0.13 | 0.23 | 0.64 |
| | | 0.25 | 0.52 | 0.96 | 0.76 | 0.95 | 0.05 | 0.05 | 0.06 | 0.23 |
| 800 | 0 | 0.05 | 1 | 1 | 1 | 1 | 0.99 | 0.70 | 1 | 1 |
| | | 0.10 | 1 | 1 | 1 | 1 | 0.87 | 0.37 | 0.90 | 1 |
| | | 0.15 | 0.89 | 1 | 0.98 | 0.99 | 0.60 | 0.29 | 0.62 | 0.83 |
| | | 0.20 | 0.61 | 0.96 | 0.75 | 0.93 | 0.13 | 0.19 | 0.24 | 0.52 |
| | | 0.25 | 0.35 | 0.72 | 0.38 | 0.69 | 0.03 | 0.14 | 0.06 | 0.16 |
| | 0.05 | 0.05 | 1 | 1 | 1 | 1 | 0.97 | 0.31 | 0.89 | 1 |
| | | 0.10 | 0.98 | 0.98 | 0.99 | 1 | 0.71 | 0.22 | 0.62 | 0.94 |
| | | 0.15 | 0.90 | 1 | 0.99 | 1 | 0.38 | 0.27 | 0.44 | 0.68 |
| | | 0.20 | 0.70 | 0.97 | 0.84 | 0.93 | 0.11 | 0.19 | 0.23 | 0.45 |
| | | 0.25 | 0.30 | 0.76 | 0.39 | 0.63 | 0.02 | 0.09 | 0.02 | 0.08 |
| | 0.10 | 0.10 | 0.97 | 0.99 | 0.99 | 1 | 0.65 | 0.07 | 0.26 | 0.90 |
| | | 0.15 | 0.97 | 1 | 1 | 1 | 0.25 | 0.03 | 0.19 | 0.57 |
| | | 0.20 | 0.78 | 0.99 | 0.89 | 0.97 | 0.08 | 0.12 | 0.12 | 0.29 |
| | | 0.25 | 0.49 | 0.90 | 0.57 | 0.85 | 0.01 | 0.02 | 0.02 | 0.08 |

*L*: sequence length, *mr*: mutation rate, and $\Delta mr$: difference in mutation rates

**Table 3. Comparison of prediction accuracy between MEME (M), WCONSENSUS (C), Weeder (W) and dMotifGreedy (D) when 30% noise was added to the datasets.**

| $w$ | $\Delta mr$ | Average $mr$ | Prediction accuracy ||||||||
|---|---|---|---|---|---|---|---|---|---|---|
| | | | ≥1 motif identified |||| 2 motifs identified ||||
| | | | M | C | W | D | M | C | W | D |
| 12, 12 | 0 | 0.05 | 1 | 1 | 1 | 1 | 1 | 0.50 | 1 | 1 |
| | | 0.10 | 0.99 | 1 | 1 | 1 | 0.79 | 0.26 | 0.97 | 1 |
| | | 0.15 | 0.96 | 1 | 1 | 1 | 0.61 | 0.22 | 0.75 | 1 |
| | | 0.20 | 0.82 | 1 | 0.98 | 1 | 0.19 | 0.16 | 0.59 | 0.88 |
| | | 0.25 | 0.56 | 0.97 | 0.86 | 0.97 | 0.10 | 0.23 | 0.33 | 0.69 |
| | 0.05 | 0.05 | 1 | 1 | 1 | 1 | 1 | 0.17 | 0.86 | 1 |
| | | 0.10 | 1 | 1 | 1 | 1 | 0.79 | 0.17 | 0.77 | 1 |
| | | 0.15 | 0.97 | 1 | 1 | 1 | 0.55 | 0.14 | 0.61 | 0.98 |
| | | 0.20 | 0.82 | 1 | 1 | 1 | 0.26 | 0.19 | 0.49 | 0.83 |
| | | 0.25 | 0.57 | 0.91 | 0.84 | 0.96 | 0.10 | 0.08 | 0.25 | 0.47 |
| | 0.10 | 0.10 | 1 | 1 | 1 | 1 | 0.71 | 0.02 | 0.20 | 1 |
| | | 0.15 | 0.99 | 1 | 1 | 1 | 0.38 | 0.10 | 0.38 | 0.94 |
| | | 0.20 | 0.88 | 1 | 1 | 1 | 0.15 | 0.05 | 0.29 | 0.68 |
| | | 0.25 | 0.68 | 0.96 | 0.96 | 1 | 0.08 | 0.06 | 0.07 | 0.34 |

$w$: motif width, $mr$: mutation rate, and $\Delta mr$: difference in mutation rates

**Table 4. Performance comparison between MEME (M), WCONSENSUS (C), Weeder (W) and dMotifGreedy (D) based on pairs of real motifs.**

| Organism | Number of motif pairs | Number of successful identification ||||||||
|---|---|---|---|---|---|---|---|---|---|
| | | ≥1 motif identified |||| 2 motifs identified ||||
| | | M | C | W | D | M | C | W | D |
| E. coli K12 | 24 | 18 | 23 | 16 | 22 | 10 | 7 | 5 | 11 |
| Human | 18 | 16 | 18 | 13 | 17 | 11 | 2 | 5 | 12 |
| Drosophila | 33 | 10 | 32 | 32 | 29 | 2 | 19 | 12 | 20 |